\documentclass[12pt]{article}

\usepackage{amsmath,amssymb}
\usepackage{graphicx}

\makeatletter

\renewcommand\section{\@startsection{section}{1}{\z@}
                                   {-3.5ex \@plus -1ex \@minus -.2ex}
                                   {2.3ex \@plus .2ex}
                                   {\normalfont\large\bfseries}}
\renewcommand\subsection{\@startsection{subsection}{2}{\z@}
                                   {-3.25ex\@plus -1ex \@minus -.2ex}
                                   {1.5ex \@plus .2ex}
                                   {\normalfont\normalsize\bfseries}}
\renewcommand\subsubsection{\@startsection{subsubsection}{3}{\z@}
                                   {-3.25ex\@plus -1ex \@minus -.2ex}
                                   {1.5ex \@plus .2ex}
                                   {\normalfont\normalsize\bfseries}}
\renewcommand\paragraph{\@startsection{paragraph}{4}{\z@}
                                   {3.25ex \@plus1ex \@minus.2ex}
                                   {-1em}
                                   {\normalfont\normalsize\bfseries}}

\makeatother

\newcommand{\be}{\begin{equation}}
\newcommand{\ee}{\end{equation}}
\newcommand{\bea}{\begin{eqnarray}}
\newcommand{\eea}{\end{eqnarray}}
\newcommand{\ba}{\begin{array}}
\newcommand{\ea}{\end{array}}

\newcommand{\id}{\hbox{1\kern-.27em l}}
\newcommand{\lb}{\langle}
\newcommand{\rb}{\rangle}
\newcommand{\half}{ {\textstyle \frac{1}{2}  } }

\newcommand{\al}{\alpha}
\newcommand{\ga}{\gamma}
\newcommand{\Ga}{\Gamma}
\newcommand{\bet}{\beta}

\newcommand{\ka}{\kappa}
\newcommand{\de}{\delta}

\newcommand{\ep}{\epsilon}
\newcommand{\si}{\sigma}
\newcommand{\la}{\lambda}

\newcommand{\Om}{\Omega}
\newcommand{\ze}{\zeta}

\newcommand{\La}{\Lambda}

\newcommand{\tha}{\theta}

\newcommand{\Ups}{\Upsilon}

\newcommand{\dal}{\dot{a}}
\newcommand{\dbeta}{\dot{b}}

\newcommand{\ua}{\underline{a}}
\newcommand{\ub}{\underline{b}}
\newcommand{\uc}{\underline{c}}
\newcommand{\ud}{\underline{d}}

\newcommand{\bA}{\bar{A}}

\newcommand{\bW}{\bar{W}}
\newcommand{\bd}{\bar{d}}
\newcommand{\bp}{\bar{p}}

\newcommand{\btha}{\bar{\theta}}
\newcommand{\bla}{\bar{\lambda}}

\newcommand{\cN}{\mathcal{N}}

\newcommand{\D}{{\rm d}}
\newcommand{\pa}{\partial}
\newcommand{\rar}{\rightarrow}
\newcommand{\non}{\nonumber}

\newcommand{\SU}{\mathrm{SU}}
\newcommand{\SO}{\mathrm{SO}}
\newcommand{\Sp}{\mathrm{Sp}}
\newcommand{\U}{\mathrm{U}}

\newcommand{\ts}{\textstyle}

\newcommand{\tal}{\widetilde{\alpha}}

\newcommand{\tla}{\widetilde{\lambda}}

\newcommand{\behat}{\beta'}

\newcommand{\gahat}{\gamma'}

\newcommand{\chihat}{\chi'}
\newcommand{\phihat}{\phi'}

\begin{document}

\setcounter{footnote}{0}
\stepcounter{table}

{\small
\begin{flushright}
CERN-PH-TH/2005-170\\
\end{flushright}
}

\vspace{-2mm}

\begin{center}
{\Large\sf Pure-spinor superstrings in $d=2,4,6$}

\vskip 5mm

Niclas Wyllard 

\vskip 3mm
Department of Physics, Theory Division, CERN, \\
1211 Geneva 23, Switzerland

\vskip 3mm 

{\tt wyllard@cern.ch}
\end{center}
\begin{abstract}
\noindent We continue the study of the $d=2,4,6$  
pure-spinor superstring models introduced in \cite{Grassi:2005}. 
By explicitly solving the pure-spinor constraint we show that these theories 
 have vanishing central charge and work out the (covariant) current algebra 
for the Lorentz currents. 
We argue that these super-Poincar\'e covariant models may be thought of 
as compactifications of the superstring on $CY_{4,3,2}$, and take some 
steps toward making this precise by constructing a map to the RNS superstring 
variables. We also discuss the relation to the so called hybrid superstrings, 
which describe the same type of compactifications. 
\end{abstract}

\setcounter{equation}{0}
\section{Introduction}
A couple of years ago Berkovits proposed a new approach to the quantisation 
of the ten-dimensional superstring \cite{Berkovits:2000a} 
(see also \cite{Grassi:2001} for variants  
of this idea). This so called pure-spinor superstring has the virtue 
that it has manifest ten-dimensional super-Poincar\'e covariance. 
For a review see \cite{Berkovits:2002d}.

A natural question to ask is if there are pure-spinor superstrings in 
lower dimensions, for instance arising via compactifications.
In \cite{Grassi:2005} we introduced pure-spinor 
superstring theories\footnote{Pure-spinor superparticle 
models in  $d=4,6$ were introduced earlier in section 2.6 
of~\cite{Berkovits:2002d} and  correspond to the particle limit of the 
models in \cite{Grassi:2005}.} in $d=2,4,6$ by mimicking Berkovits' 
construction in $d=10$ \cite{Berkovits:2000a}.
As in $d=10$, the ``ghost'' sector of these models involve    
constrained (``pure'')  bosonic spinors. 
The (quadratic) constraints on these spinors were first written down 
in \cite{Berkovits:2002d} and were discussed in \cite{Grassi:2005}.

In this paper we show that the constraints on the $\la$'s are such 
that they imply that the world-sheet conformal field theories 
for the $d=2,4,6$ models have $c=0$ (vanishing total central charge) 
and $k=1$ (total matter+ghost Lorentz current algebra has level one). 
The approach we follow to obtain these results is the same as the 
one originally followed in $d=10$~\cite{Berkovits:2000a,Berkovits:2001a}.  
This is perhaps not the most elegant method since it temporarily 
breaks manifest covariance\footnote{As in $d=10$, in $d=2n$ we solve the 
pure-spinor constraint in terms of free fields by temporarily breaking 
the manifest $\SO(2n)$ (Wick rotated) Lorentz invariance to 
$\U(n) \simeq \SU(n) {\times} \U(1).$}, but on the other hand 
the method is straightforward and leads to expressions 
which can be compared with corresponding ones in the Ramond-Neveu-Schwarz 
(RNS) superstring. For an alternative covariant approach 
see~\cite{Berkovits:2005,Grassi:2005b}.

In view of the well-known fact that the superstring is only consistent 
in $d=10$ the above results might seem surprising. 
The existence of $d=4,6$ pure-spinor {\it superparticle} models is 
not surprising given that covariant Brink-Schwarz superparticle 
models~\cite{Brink:1981} exist in these dimensions. 
The $d=4,6$ Brink-Schwarz superparticle models can be quantised 
in the light-cone gauge and it should be possible to relate them to the 
pure-spinor models (this appears to be slightly subtle though; see section 2.6 
in \cite{Berkovits:2002d} for a discussion).
On the other hand, if one tries to quantise the corresponding 
$d=4,6$ Green-Schwarz (GS) superstring theories \cite{Green:1983} in the 
light-cone gauge one finds that this leads to an inconsistency since the 
Lorentz algebra does not close (see e.g.~\cite{Green:1987} for a discussion).
For the pure-spinor models, in contrast to the situation in 
the light-cone GS superstring, Lorentz covariance is not 
broken by quantum effects. However, we do {\em not} claim that the pure-spinor 
models are consistent critical superstring theories\footnote{For a recent 
discussion of covariant non-critical superstrings see \cite{Grassi:2005a}.} 
in dimensions $d<10$. 
Rather we will argue that the pure-spinor models should be thought 
of as the non-compact piece of compactifications 
of the ten-dimensional superstring on 4,3,2 (complex) dimensional 
Calabi-Yau (CY) manifolds from $d=10$ to $d=2,4,6$. 
The inconsistency of the lower-dimensional theory without 
the CY piece, which in the RNS superstring appears as $c\neq 0$ and 
in the light-cone GS superstring as a breakdown of Lorentz covariance, 
for the pure-spinor models should appear in another sector of the 
theory (exactly where the inconsistency appears we do not yet 
fully understand).

An important question to understand is the relation between the $d=2,4,6$ 
pure-spinor models and the compactified RNS superstring. 
We will argue that the relation is 
analogous to the one in the uncompactified case \cite{Berkovits:2001a} 
(see also \cite{Aisaka:2003}). 
More precisely, we will show that after a change of variables from the 
RNS variables plus the addition of a certain number of $c=0$ 
``topological quartets'', the field content and stress tensor  precisely 
matches that of the pure-spinor models plus a decoupled sector describing 
the CFT of the compactification manifold.

With the interpretation of the pure-spinor models as compactified theories 
with manifest Lorentz covariance in the non-compact directions, the question 
arises how these models are related to the so called hybrid 
superstrings \cite{Berkovits:1994a,Berkovits:1994b} which describe the same 
type of compactifications. Roughly, we find that after the addition 
of a certain $c=0$ piece, the stress tensor in the hybrid 
superstring agrees with the stress tensor of the corresponding pure-spinor 
superstring written in terms of Lorentz covariant variables. 
In comparison to the hybrid superstrings, an advantageous feature of 
the pure-spinor models is that they seemingly circumvent many of 
the problematic aspects associated with the negative-energy chiral 
scalars present in the hybrid models. 
However, it should be stressed that the pure-spinor models 
need to be developed further before they can be considered as a viable 
alternatives to the hybrid models. In particular, the BRST operator, vertex operators and scattering amplitudes need to be better understood. 
Another application of the $d=2,4,6$ pure-spinor conformal field 
theories presented in this paper is as toy models for the more 
involved $d=10$ pure-spinor conformal field theory.

This paper is organised as follows. In the next section 
the $d=4$ pure-spinor model is discussed. 
In section \ref{d6} the same analysis is carried out for the $d=6$ model, 
and in section \ref{d2} the $d=2$ model is briefly discussed. 
Then in section \ref{RNS} the relation to the RNS superstring is discussed 
followed in section \ref{hybrid} by a discussion of the relation to the 
hybrid superstrings. A discussion of open problems and some applications 
are presented in section \ref{sDisc}. Finally, in the appendix some 
technical results are collected.

\section{The $d=4$ pure-spinor superstring}
\label{d4}
In this section we consider the $d=4$, $\cN=1$ pure-spinor 
superstring \cite{Grassi:2005}. 
For simplicity we consider the type II string in a flat supergravity 
background and write only the left-moving worldsheet fields explicitly. 
The left-moving (holomorphic) ``matter'' worldsheet fields 
are  $(x^{m}, \theta^\al,p_\al)$, where $\theta^\al$ is a four-component 
Dirac spinor\footnote{The slightly unfortunate Dirac spinor index notation 
is chosen to make the $d=4$ formul\ae{} correspond as closely as possible 
to the $d=10$ ones.}, and 
$p_\al$ is its conjugate momentum ($\al=1,\ldots4$). 
The Dirac spinor $\tha^\al$ can be decomposed 
into a Weyl spinor, $\tha^a$ ($a=1,2$), and an anti-Weyl spinor, 
$\btha^{\dot{a}}$ ($\dot{a}=1,2$). 
Similarly, $p_\al$ can be decomposed into $p_a$, and $\bp_{\dot{a}}$.
The free fields $x^m$, $\tha^\al$ and $p_\al$ have the standard OPEs 
(in units where $\alpha'=2$)
\be
x^{m} (y, \bar{y}) \, x^{n} (z, \bar{z}) \sim -  \eta^{mn} \log
\left| y - z \right|^{2} \,, \qquad  p_{\alpha} (y) \, \theta^{\beta} (z) \sim
\frac{\delta_{\alpha}^{\beta}}{y-z} \,.
\ee
In the Weyl basis one has
\be
 p_{a} (y) \, \theta^{b} (z) \sim
\frac{\delta_{a}^{b}}{y-z} \,, \qquad 
\bp_{\dal} (y) \, \btha^{\dbeta} (z) \sim
\frac{\delta_{\dal}^{\dbeta}}{y-z}\,.    
\ee
As in the $d=10$ pure-spinor formalism, the world-sheet ghost fields 
involve a Grassmann-even spinor, $\la^\al$  
(and its conjugate momentum). The bosonic (Dirac) spinor $\la^\al$ is assumed 
to satisfy the ``pure spinor'' condition 
\bea \label{4dpure-spinor}
\lambda \Gamma^{m} \lambda = 0\,.
\eea
Here $\Gamma^m_{\al\beta}$ are the (symmetric) $4{\times}4$ gamma matrices 
(we do not explicitly write the charge conjugation matrix used to lower 
indexes). In the Weyl basis, the above condition can be written as 
$\la^a \bla^{\dot{b}} = 0$ (more details are given below).

The conjugate momentum to $\la^\al$ will be denoted $w_\al$. 
As in the $d=10$ pure-spinor superstring, because of the pure-spinor 
constraint, $w_\al$ is only defined 
up to the gauge invariance: $w_\al \rar w_\al + \La_m (\Ga^m \la)_\al$.

As in $d=10$, one can construct the Lorentz covariant 
quantities 
\be \label{4dNh}
 N^{mn} = \half \, w \Gamma^{mn} \lambda \,,  \qquad 
\partial h = \half\, w \lambda \,.  
\ee
To calculate the OPEs between involving $N^{mn}$ and $\pa h$ 
one can either (temporarily) 
explicitly solve the pure-spinor constraint and calculate the OPEs 
by expressing $N^{mn}$ and $\pa h$ in terms 
of the unconstrained variables, verifying Lorentz covariance at the end, 
or one can use the covariant method introduced 
in~\cite{Berkovits:2005}. Here we will follow the former approach, which 
has the advantage that it also gives us information about the zero-mode 
saturation rule and a possible relation to the RNS and hybrid 
superstring (see later sections). The covariant approach is 
discussed in \cite{Grassi:2005b}.

One can solve the pure-spinor constraint (\ref{4dpure-spinor}) in terms of 
free fields by temporarily breaking the $\SO(4)$ (Wick rotated) 
Lorentz group to $\U(2) \simeq \SU(2) {\times} \U(1)$. 
Under this subgroup $\la^\al$ decomposes into (see appendix~\ref{un} 
for further details) $(\la^+,\la^a,\la_{ab})$, 
where $\la_{ab}=-\la_{ba}$.  Here $a,b = 1, 2$. 
In this $\U(2)$ basis 
the pure-spinor condition becomes
\be \label{4dpure-constr}
\la^+ \la^a = 0\,,  \qquad \la_{ab} \la^b=0\,. 
\ee
We have chosen to write the $d=4$ pure spinor condition 
in a way which is closely analogous to the $d=10$ case. 
The $d=4$ pure-spinor condition can also
be conveniently written as $\la^a \bla^{\dot{a}}=0$  where we have 
introduced the notation $\bla^{\dot{a}} = \{\la^+, \half \ep^{ab}\la_{ab} \} $.
In our conventions, as the notation suggests, $\la^a$ is a Weyl spinor whereas $\bla^{\dal}$ is an 
anti-Weyl spinor.

The explicit solution to the constraints (\ref{4dpure-constr}) is
\begin{equation} \label{lapar}
\{\lambda^{a} = 0 \}\cup \{\bla^{\dot{a}}=0 \} \,. 
\end{equation}
It is important to note that the solution has two ``patches'', $\la^a=0$ and 
$\bla^{\dal}=0$. Both patches are part of the solution and should be taken 
into account (see \cite{Grassi:2005} for a discussion). 
Below we are mostly concerned with quantities which do not depend on 
where in the space of solutions (\ref{lapar}) one is and we therefore 
work in one of the two patches, $\la^a=0$. 

The pure-spinor constraint (\ref{4dpure-spinor}) eliminates two components 
from $\la^\al$.
The remaining independent (unconstrained) components of $\la^\al$ 
have canonical free-field OPEs with the corresponding components 
of  the conjugate momentum.  
The components of the conjugate momentum corresponding to the two  
eliminated components of $\la^\al$ can be gauged to zero.  
For instance the gauge $w_a=0$ can be chosen in the $\la^a=0$ patch. 

In the $\U(2)$ basis the Lorentz covariant quantities (\ref{4dNh}) 
can be written
\bea \label{4dNs}
&& N^{ab} = \half w^{ab} \la^+ \,, \qquad N_{ab} = - \half w_+ \la_{ab} - 
\half \frac{\pa \la_{ab}}{\la^+} \,,  \non \\
&& N^a{}_b = -{\ts \frac{1}{4} } \de^a_b [ w_+ \la^+ + \frac{\pa \la^+}{\la^+}
- \half w^{ab} \la_{ab} ] \,,
\eea
as well as
\be \label{4dpah}
\pa h =  \half w_+ \la^+ - \half \frac{\pa \la^+}{\la^+} 
+ {\ts \frac{1}{4}} w^{ab} \la_{ab} \,.
\ee
Here the terms with derivatives are related to normal-ordering ambiguities. The normal-ordering 
ambiguities are chosen so that $\pa h$ and $N^{mn}$ have no OPEs with 
each other and $\SO(4)$ Lorentz covariance is preserved (see below).

The above ghost Lorentz currents (\ref{4dNs}) satisfy the OPEs
\bea \label{4dNNnc}
N^{a}{}_{b}(y) N^{c}{}_{d}(z) &\sim& 
\frac{k}{4} \frac{\de^a_d\de^c_b}{(y - z)^2}
+ \frac{1}{2} \frac{\de^a_d N^c{}_b - \de^c_b N^a{}_d}{y - z} \non \\
N^{ab}(y) N^{c}{}_{d}(z) &\sim& -\frac{\de^{[a}_d N^{b] c} }{y - z} \non \\
N^{ab}(y) N_{cd}(z) &\sim& -\frac{k}{2} \frac{\de^{ab}_{cd} }{(y - z)^2}
+ 2 \frac{\de^{[a}_{[c} N^{b]}{}_{d]}}{y - z}  \\
N_{ab}(y) N^{c}{}_{d}(z) &\sim&  \frac{\de_{[a}^c N_{b]d} }{y - z} \non
\eea
where $k=0$ and, as usual, 
$\de^{ab}_{cd} = \frac{1}{2}(\de^a_c\de^b_d - \de^a_d\de^b_c)$.
In manifestly $\SO(4)$ covariant notation one finds that the OPEs involving 
$N^{mn}$ and $\la^\al$ take the form 
\bea \label{NOPEs4d}
N^{mn} (y) \lambda^\al (z) &\sim& \frac{1}{2}\ \frac{1}{y - z}
\left( \Gamma^{mn} \la \right)^\al (z) \qquad \quad \;  \\[3pt]
N^{p q } (y) N^{mn} (z) &\sim& \frac{ 
\eta^{p m} N^{q  n} (z) {-} \eta^{q m} N^{p n} (z) 
- (m\leftrightarrow n)}{y - z} \,. \non
\eea
The second equation in (\ref{NOPEs4d}) shows that $N^{mn}$ satisfies 
an $\SO(4)$ current algebra with level $k=0$.  
Furthermore, $ h$ has no singular OPEs with $N^{mn}$ and satisfies
\begin{equation}\label{extra1}
h(y) h(z) \sim - \log \left( y - z \right), \qquad \partial h(y)
\lambda(z) \sim \frac{1}{2}\ \frac{1}{y-z} \lambda(z)\,.
\end{equation}
Notice that the above results are completely analogous to the ones in 
$d=10$ (see e.g.~section 2 of ref.~\cite{Schiappa:2005} for a discussion 
of the $d=10$ case using closely related notation and conventions).

Above we found that $N^{mn}$ satisfies an $\SO(4)$ current algebra with 
level $k=0$. 
In comparison, the OPEs involving the $(p,\tha)$ Lorentz currents, 
$M^{mn} = -\frac{1}{2}\, p \Gamma^{mn} \theta$, take the form
\begin{eqnarray} \label{4dOPE-M}
M^{mn} (y) \theta^{\alpha} (z) &\sim&  \frac{1}{2}\ \frac{1}{y - z}
( \Gamma^{mn} \theta)^\al(z) 
\qquad \quad \;\;  \\
M^{p q} (y) M^{mn} (z) &\sim& 
\frac{ \eta^{p m } M^{q n} (z) {-} \eta^{ q m }
M^{p  n} (z) - (m\leftrightarrow n)
}{y - z} +  \frac{\eta^{p n} \eta^{q m} {-} \eta^{p m}
\eta^{q n}}{(y - z)^{2}} \,. \non
\end{eqnarray}
Thus the $M^{mn}$'s form an $\SO(4)$ current
algebra with level $k=1$. The total Lorentz current,  
$L_{mn} = M^{mn} + N^{mn}$, thus satisfies the OPE
\begin{eqnarray}
\!\!\!\!\!\!\!\!\!\! 
L^{p q} (y) L^{mn} (z) \!\!\! &\sim& \!\!\! \frac{ \eta^{p m}
L^{q n}(z)  {-} \eta^{q m}
L^{p  n}(z) - (m\leftrightarrow n) }{y - z} 
+ \frac{\eta^{p n} 
\eta^{q m} {-} \eta^{p m}
\eta^{q n}}{(y - z)^{2}}\,,
\label{4dOPE-LL}
\end{eqnarray}
forming a current algebra with level $k=1$ (just as in the 
$d=10$ case).

The ``matter'' part of the stress tensor for 
the $d=4$ pure-spinor superstring can be written \cite{Grassi:2005}
\be \label{4dTmat}
T_{\rm mat} = -\half \pa x_m \pa x^m - p_\al \pa \theta^\al \,.
\ee
From this expression one sees that the $x^m$ CFT has 
central charge $c=4$, while the $(p,\theta)$ 
CFT has central charge $c=-8$. In order for the total 
central charge to vanish, the ghost CFT has to have $c=4$. 

Using the covariant fields $N^{mn}$ and $\pa h$, the 
 $d=4$ ghost stress tensor can be written in a manifestly 
Lorentz invariant way  as 
\be\label{covst4N}
T_{N,\pa h} 
= - {\ts \frac{1}{8}} N_{mn} N^{mn} - \half \left
( \partial h \right)^{2} + \half \partial^{2} h  \,.
\ee
The terms in the above expression are all that are allowed by Lorentz 
invariance and the conformal weight of the stress tensor. 
The requirement that $N^{mn}$ should have conformal weight 
one as should $\pa h$ (except for the background 
charge) and that $\la$ should have conformal weight zero puts restrictions 
on the coefficients.

By using (\ref{4dNs}) and (\ref{4dpah}) together with various normal ordering 
rearrangements, one can check that the stress tensor (\ref{covst4N}) 
reduces to 
\begin{equation}
T_{w,\la} =  w_+  \partial \la^{+} + \half w^{ab}\pa \la_{ab} \,, 
\end{equation}
\noindent
written in terms of the $\U(2)$ variables. Thus the  ghost sector 
comprise two $\beta\gamma$ systems of weight one. The central charge 
of the ghost stress tensor is therefore $c=4$ and hence the total central 
charge vanishes. This result 
can also be seen directly from (\ref{covst4N}). The first piece
involves the ghost Lorentz currents, $N^{mn}$, and is a Sugawara
construction for an $\SO(4)$ WZNW  model with level $k=0$. 
Indeed, recalling that the dual Coxeter number of $\SO(2n)$ is 
$g^\vee=2n-2$, we find\footnote{The prefactor in front of 
$N_{mn}N^{mn}$  in (\ref{covst4N}) is $-\frac{1}{4(k+g^\vee)}$. 
To obtain the more conventional 
$+\frac{1}{2(k+g^\vee)}$ one would have to rescale the currents $N^{mn}$.} 
$2(g^\vee + k) = 4$. Using standard formul\ae{},  
the central charge for an $SO(2n)$ Lorentz current algebra with level $k$ is
\begin{equation}
c = \frac{k \dim \SO(2n)}{ k+g^\vee}  \,,
\end{equation}
which vanishes when $k=0$. In (\ref{covst4N}) the pieces involving $\pa h$ 
refer to a Coulomb gas with background charge $Q=1$, 
and consequently central charge $c = 1 + 3 Q^{2} = 4$. 
As above, the total ghost central charge is $c=4$. 

As in $d=10$ it turns out that one can also write the 
$(p, \tha)$ part of the stress tensor 
in a form similar to what was done above for the ghost part. 
It is unclear if this is just a curiosity or whether it 
can be useful. 
The analogue of $\pa h$ in the $(p,\theta)$ sector is  
$\pa g  = \frac{1}{2} \, p_{\al} \tha^\al$. 
One can also introduce  
$\pa \hat{g} = \frac{1}{2} \, p \Ga_5 \tha $. 
 The two scalars $g$, $\hat{g}$ have no singularities 
with $M^{mn}$ and satisfy
\be
g (y) \, g(z) \sim  
 \log(y-z) \,, \qquad \hat{g} (y) \, \hat{g}(z) \sim  
 \log(y-z) \,, \qquad g(y) \, \hat{g}(z) \sim  0\,.
\ee
In terms of $M_{mn}$, $\pa g$ and $\pa \hat{g}$ one finds that 
$T_{p\tha} = - p_{\al} \pa \tha^\al$ can be rewritten as
\be \label{covst4M}
T_{M,\pa g} 
= -{\ts \frac{1}{12}} M_{mn} M^{mn} 
+ \half \left (\partial \hat{g} \right)^{2} 
+ \half \left (\partial g \right)^{2} 
- \, \partial^{2} g \,.
\ee
The central charge obtained from (\ref{covst4M}) is 
$c = 2 + 1 + 1 - 3\cdot 2^2 = -8$ as it should be. 

In contrast to the $d=10$ case there is a structural difference 
between (\ref{covst4N}) and (\ref{covst4M}) in that in the latter 
$\hat{g}$ appears but in the former no corresponding $\hat{h}$ appears. 
As we will see in the next section a similar result also holds for the 
$d=6$ model (see this section for a longer discussion). 

The zero-mode saturation rule for the above model can be obtained by 
noting that $T_{p,\tha}$ comprise four $bc$-type systems 
and $T_{w,\la}$ comprise two $\beta\ga$-type 
systems, all of weight one. By using standard methods 
\cite{Friedan:1985} one finds that the saturation rule is
\be \label{cansat4}
\lb 0| \ep_{\dot{a}\dot{b}}\btha^{\dot{a}}\btha^{\dot{b}} 
[\tha^{+} \tha_{ab} ]
[\de(\la^+) \de(\la_{ab}) ] |0 \rb \neq 0  \,,
\ee
which is equivalent to
\be
\lb 0| \tha^2 |\Omega \rb \neq 0 \,,
\ee
where (as in $d=10$) $|\Omega \rb = \prod_{A=1}^2 Y^A |0\rb$ with 
$Y^A = C^A_\al \tha^\al \de(C^A_\al \la^\al)$ and $C^A_\al$ are certain 
constant bosonic spinors.
The two operators $Y^A$ each carry $(\la,\tha)$ charge $(-1,+1)$ and 
one can check that the above saturation rule is consistent with the background 
charges of $\pa h$ and $\pa g$. 

The next step in the analysis of the above model would be to construct 
a BRST operator and analyse vertex operators and scattering amplitudes. 
These questions were touched upon in \cite{Grassi:2005}. The naive BRST 
operator (based on the $d=10$ expression)
\begin{equation} \label{4dQBRST}
Q = \oint \lambda^{\alpha} d_{\alpha}\,, 
\end{equation}
has {\it off-shell} $\cN=1$ super-Yang-Mills as its massless 
cohomology at ghost number 1 \cite{Berkovits:2002d,Grassi:2005}.  
The fact that the massless cohomology is off-shell SYM seems to 
indicate that the above BRST operator can not be the full story. 
Vertex operators corresponding to the above BRST operator were 
briefly discussed in \cite{Grassi:2005}. The fact that the unintegrated 
vertex operator for the massless states, $U$, has ghost number 1 together 
with the form of the above saturation rule seems to require that the three 
unintegrated vertex operators in tree-level scattering amplitudes should 
have total ghost number zero. 
This seems to indicate that the same construction as 
in $d=10$~\cite{Berkovits:2000a,Berkovits:2004a} can not work here 
without modification. We hope to return to these questions in the future.

To summarise: in this section we analysed the conformal field theory for the 
$d=4$ pure-spinor superstring and showed that it has vanishing central charge 
and is such that the Lorentz current algebra has level one. 
We also obtained the zero-mode saturation rule. Many open problems remain, 
some of which were mentioned above.

\section{The $d=6$ pure-spinor superstring}
\label{d6}
In this section the same analysis that was carried out in $d=4$ in 
the previous section will be performed for the $d=6$ case with minimal 
supersymmetry. As in $d=4$ we work in a flat type II supergravity background 
and only display the left-moving sector. 
The left--moving (holomorphic) ``matter'' 
worldsheet fields in $d{=}6$ with $\cN=(1,0)$ supersymmetry 
are \cite{Grassi:2005} $(x^{m}, \theta^\al_I,p_\al^I)$, where $\theta^\al_I$ 
is a doublet ($I=1,2$) of four--component Weyl spinors, and $p_\al^I$ 
are their conjugate momenta. (Note that in this section $\al,\beta,\ldots$ 
denote Weyl indexes, and not Dirac indexes as in $d=4$.)

By analogy with the $d=10$ and $d=4$ cases we take the  
world--sheet ghost fields to involve a doublet of Grassmann-even 
Weyl spinors, $\la^\al_I$ ($I=1,2$), (and their conjugate momenta) and 
impose the pure-spinor condition~\cite{Berkovits:2002d,Grassi:2005}
\be \label{d6pure-spinor}
\ep^{IJ} \lambda_I \gamma^{m} \lambda_J = 0 \,.  
\ee
Here $\gamma^{m}$ are the (antisymmetric) $4 {\times} 4$ 
off--diagonal blocks (``Pauli matrices'') in the Weyl representation 
of the $8 {\times} 8$ six--dimensional gamma matrices $\Gamma^{m}$. 
Note that the above condition (\ref{d6pure-spinor}) is {\it not} a 
conventional pure-spinor condition in the sense of Cartan 
(which is solved by a Weyl spinor). However, as confusion is unlikely 
to arise we refer to (\ref{d6pure-spinor}) as a pure-spinor 
condition throughout.

The free fields $x^m$, $\tha^\al_I$ and $p_\al^I$ have the standard OPEs 
(in units where $\alpha'=2$),
\be x^{m} (y, \bar{y}) x^{n} (z, \bar{z}) \sim -  \eta^{mn} \log
\left| y - z \right|^{2} \,, \qquad
  p^I_{\alpha} (y) \theta_J^{\beta} (z) \sim
\frac{\delta^I_J\delta_{\alpha}^{\beta}}{y-z}  \,.
\ee
As in $d=4,10$ we can solve the pure-spinor constraint in terms of 
free fields by temporarily breaking the manifest $\SO(6)$ (Wick rotated) 
Lorentz invariance to $\U(3) \simeq \SU(3) {\times} \U(1)$. 
Under this subgroup $\la_I^\al$ decomposes into 
(see appendix~\ref{un} for further details) 
$(\la_I^+,\la_I^a)$, where $a,b = 1, \ldots, 3$.

In the $\U(3)$-basis the pure-spinor condition becomes
\be \label{d6pure-constr}
\ep^{IJ}\la^+_I \la^a_J=0\,, 
 \qquad \ep_{abc} \ep^{IJ} \la^a_I \la^b_J =0 \,.
\ee
An explicit solution to these constraints is
\begin{equation} \label{d6lapar}
\la_2^a = \frac{\la^+_2}{\la^+_1}\la_1^a \,,  
\end{equation}
which can also be written as $\la_2^\al = e^{-v} \la_1^\al$, 
where $v = \ln(\la_1^+/\la_2^+)$.
Thus, we see that the pure-spinor condition (\ref{d6pure-spinor}) 
eliminates $3$ components from $\la^\al_I$. 
The remaining five independent (unconstrained) components of $\la_I^\al$ 
have canonical free-field OPEs with the corresponding components 
of the conjugate momenta, $w_\al^I$. 
Using the gauge symmetry induced by the pure-spinor constraint, 
$\de w^I_\al = \ep^{IJ}\La_m (\ga^m\la_J)_\al$ 
the components of the conjugate momentum corresponding to the three  
constrained components of $\la_I^\al$ can be gauged to zero, 
e.g.~$w^2_a=0$. 

As in $d=4,10$, one can construct the
 $\SO(6)$ Lorentz covariant quantities 
\be
 N^{mn} = \half w^I \gamma^{mn} \lambda_I \,, \qquad  
\partial h = \half w^I \lambda_I \,.
\ee
In the $\U(3)$ basis one has:
\bea \label{6dNs}
&& N^{ab} = \half \ep^{abc} w^1_c \la_1^+ \,, \non \\
&&N_{ab} = - \half \ep_{abc} w^1_+ \la_1^{c} 
- \half \ep_{abc} (w^2_+\la^+_2) \frac{\la_1^{c}}{\la_1^+}
+ {\ts \frac{a_1}{2}} \ep_{abc} \frac{\pa \la_1^+}{(\la_1^+)^2} \la^c_1
+ {\ts \frac{-a_1-\half}{2}} \ep_{abc} \frac{\pa \la_1^+}{\la_1^+\la^+_2} 
\la^c_1 \,, \non \\
&& N^a{}_b = -\half w_b^1 \la_1^a -{\ts \frac{1}{4} } \de^a_b 
[ w^1_+ \la_1^+ 
+ w_+^2 \la^+_2 -  w_{c}^1 \la^{c}_1 - a_1 \frac{\pa \la_1^+}{\la_1^+} 
+ (a_1+\half) \frac{\pa \la_2^+}{\la_2^+} ] ,
\eea
and
\be \label{6dpah}
\pa h = \half [ w^1_+ \la_1^+ 
+ w_+^2 \la^+_2 +  w_{c}^1 \la^{c}_1 + (2-a_1) \frac{\pa \la_1^+}{\la_1^+} 
+ (a_1 - {\ts \frac{3}{2}}) \frac{\pa \la_2^+}{\la_2^+} ] \,.
\ee
It will also be convenient to introduce 
\bea \label{6duv}
\pa u &=& \half [w_+\la^+_1 + w_c^1 \la^c_1 - w_+^2 \la^+_2 
+ (a_1+1)\frac{\pa \la^+_1}{\la_1^+} 
+ (a_1 - {\ts \frac{3}{2}}) \frac{\pa \la^+_2}{\la_2^+} ] \,, \non \\
\pa v &=& \frac{\pa \la^+_1}{\la_1^+} - \frac{\pa \la^+_2}{\la_2^+} \,.
\eea
In the above expressions, the terms with derivatives are related 
to normal-ordering ambiguities and $a_1$ is an arbitrary constant. 
We have indicated the normal-ordering prescription 
by parentheses. The normal-ordering terms are restricted by the requirement 
that the OPEs be Lorentz covariant, and the fact that they can be chosen 
 such that this is true is a non-trivial result.

The above Lorentz currents for the ghosts satisfy, in the 
$\U(3)$ basis (\ref{6dNs}), 
the same OPEs as in (\ref{4dNNnc}) but with $k=-1$. 
In manifestly $\SO(6)$ covariant notation the OPEs involving 
$N^{mn}$ and $\la_I^\al$ take the form 
\bea \label{6dNNOPE}
N^{mn} (y) \lambda_I^{\alpha} (z) &\sim &\frac{1}{2}\ \frac{1}{y - z}
{\left( \gamma^{mn} \right)^{\alpha}}_{\beta} \, \lambda_I^{\beta} (z) \,,
\\[3pt] N^{p q } (y) N^{mn} (z) &\sim& \frac{ 
\eta^{p m} N^{q  n} (z) {-} \eta^{q m} N^{p n} (z) 
- (m\leftrightarrow n)}{y - z}  -
 \frac{\eta^{p n} \eta^{q m} {-} \eta^{p m}
\eta^{q n}}{(y - z)^{2}} \,. \non
\eea
Thus, the $N^{mn}$'s form an $\SO(6)$ current algebra with level $k=-1$.
As in $d=4,10$ $h$ has no singular OPEs with $N^{mn}$, and 
satisfies
\begin{equation} \label{d6hh}
h(y) h(z) \sim - \log \left( y - z \right), \quad \partial h(y)
\lambda^\al_I(z) \sim \frac{1}{2}\ \frac{1}{y-z} \lambda^\al_I(z)\,.
\end{equation}
In addition, the worldsheet fields $\pa u$ and $\pa v$ have 
no singularities with $N^{mn}$ or $\pa h$ or with themselves 
and satisfy 
\be
\pa u(y) \, \pa v(z) \sim \frac{1}{(y-z)^2} \,.
\ee
Furthermore,
\be
\pa u(y) \,  \la_1^\al(z) \sim \frac{1}{2} \frac{1}{(y-z)} \la^\al_1 \,,
\qquad
\pa u(y) \,  \la_2^\al(z) \sim -\frac{1}{2} \frac{1}{(y-z)} \la^\al_2 \,.
\ee

In comparison to the results in (\ref{6dNNOPE}) the OPEs involving 
the $(p,\tha)$ Lorentz currents, 
$M^{mn} = -\frac{1}{2} p^I \gamma^{mn} \theta_I$, take the form
\begin{eqnarray} \label{OPE-M}
M^{mn} (y) \theta_I^{\alpha} (z) &\sim&  \frac{1}{2}\ \frac{1}{y - z}
{\left( \gamma^{mn} \right)^{\alpha}}_{\beta} \, \theta_I^{\beta} (z) \\
M^{p q} (y) M^{mn} (z) &\sim& 
\frac{ \eta^{p m } M^{q n} (z) {-} \eta^{ q m }
M^{p  n} (z) - (m\leftrightarrow n)
}{y - z} +
2 \frac{\eta^{p n} \eta^{q m} {-} \eta^{p m}
\eta^{q n}}{(y - z)^{2}} \,. \non
\end{eqnarray}
Thus the $M^{mn}$'s form an $\SO(6)$ current
algebra at level $k=2$. The total Lorentz current 
$L_{mn} = M^{mn} + N^{mn}$ satisfies the OPE
\begin{eqnarray}
\!\!\!\!\!\!\!\!\!\! 
L^{p q} (y) L^{mn} (z) \!\!\! &\sim& \!\!\! \frac{ \eta^{p m}
L^{q n}(z)  {-} \eta^{q m}
L^{p  n}(z) - (m\leftrightarrow n) }{y - z} 
+ \frac{\eta^{p n} 
\eta^{q m} {-} \eta^{p m}
\eta^{q n}}{(y - z)^{2}} \,,
\label{OPE-LL}
\end{eqnarray}
and thus forms a current algebra with level $k=1$.

The ``matter'' part of the stress tensor for 
the $d=6$ pure-spinor superstring can be written
\be
T_{\rm mat} = -\half \pa x_m \pa x^m - p^I_\al \pa \theta_I^\al \,.
\ee
From this expression one sees that the $x^m$ CFT has 
central charge $c=6$, while the $(p,\theta)$ 
CFT has central charge $c=-16$. In order for the total 
central charge to vanish, the ghost CFT has to have $c=10$. 

Using the Lorentz covariant fields $N^{mn}$, $\pa h$, $\pa u$ and $\pa v$ 
the ghost stress tensor can be written in a manifestly Lorentz invariant way 
as 
\be \label{covst6}
T_{N,\pa h} 
= - {\ts \frac{1}{12}} N_{mn} N^{mn} - \half \left
 (\partial h \right)^{2} +  \partial^{2} h + \pa u \pa v - \pa^2 v \,.
\ee
After using various normal-ordering rearrangements it can be shown that 
the above stress tensor reduces to
\be \label{6dwl}
T_{w,\la} = [(w_+^2\la^+_2) + \half \frac{\pa \la_2^+}{\la_2^+}]
\frac{\pa \la_2^+}{\la_2^+} + w_+^1 \pa \la^+_1 + w^1_c \pa \la^c_1 \,.
\ee
In this form it is easy to verify the conformal dimensions of $N^{mn}$ 
and $\la^\al_I$ and check that $c=10$ so that the total central charge 
vanishes.

We note in passing that there is in fact a slight ambiguity 
in writing (\ref{covst6}) since if one uses $-\half \pa^2 v$ 
instead of $- \pa^2v$ one is lead to the same expression as 
in (\ref{6dwl}) but with the roles of $\la_1^+$ and $\la^+_2$ interchanged.

The central charge can of course also be obtained from (\ref{covst6}). 
The first piece is a Sugawara construction for a $\SO(6)$ WZNW model with 
level $k=-1$ and consequently central charge 
$c=-5$. The piece involving $\pa h$ has $c=1+3\cdot 2^2 = 13$. Finally, 
the $\pa u$, $\pa v$ piece has $c=2$ for a total of $c=10$.

One can also rewrite the $(p, \tha)$ part of the stress tensor 
in a form similar to what was done above for the ghost part. Besides 
$M^{mn} = - \half p_I \ga^{mn} \tha^I$ it is also convenient to introduce 
$\pa g^I{}_J = p^I \tha_J$, which can be decomposed into:
\be
\pa g = \frac{1}{2\sqrt{2}} \, \pa g^I{}_I\,, \quad \qquad R^I{}_J = 
\pa g^I{}_J - \half \, \de^I_J \, \pa g \,.
\ee
The fields $\pa g^I{}_J$ have no singularities with $M^{mn}$ and satisfy
\be
\pa g^I{}_J(y) \, \pa g^K{}_L(z) \sim  
4 \frac{\de^I_L\de^K_J}{(y-z)^2} + 
\frac{\delta^I_L \pa g^K{}_J -  \delta^I_L \pa g^K{}_J}{y-z}\,.
\ee
From this result it follows that $\pa g(y) \pa g(z) \sim \log(z-y)$ and 
that $R^I{}_J$ form an $\SU(2)$ (or $\Sp(2)$) current algebra 
with level $k = 4$ (which, as opposed to the $M^{mn}$ and $N^{mn}$ 
current algebras, is conventionally normalised).

One can show that $T_{p\tha} = -p^I_\al \pa \tha^\al_I$ 
is equal to 
\be \label{covst6M}
T_{M,\pa g} 
= -{\ts \frac{1}{24}} M_{mn} M^{mn} + {\ts \frac{1}{12}} R^I{}_J R^J{}_I 
+ \half \left (\partial g \right)^{2} - \sqrt{2} \, \partial^{2} g \,.
\ee
In this form, the total central charge is calculated to be  
$c = 5 + 2 + 1-3\cdot 8 = -16$ as it should. 

Note that the analogue of the $\pa g^I{}_J$'s in the ghost sector, 
$\partial h^I{}_J = \half w^I \lambda_J$, are invariant under 
the gauge transformation $\de w^I = \La^m (\ga_m \la^I)$ and 
are thus a priori allowed operators. However, not all $h^I{}_J$'s appear  
in the stress tensor (modulo normal ordering $\pa h = h^1{}_1 + h^2{}_2$ 
and $\pa u = h^1{}_1 - h^2{}_2$). This situation is similar to the $d=4$ case 
(see the previous section) where $\hat{h}$ did not appear in $T$ but 
$\hat{g}$ did. 
The fact that not all objects which are (classically) invariant under the 
$\de w^I_\al$ gauge symmetry appear in $T$ is 
perhaps not so strange, but more puzzling is the fact that we have not 
been able to choose the normal-ordering constants in such a way that both 
$N^{mn}$ and {\it all} $\pa h^I{}_J$'s satisfy Lorentz-covariant 
{\it and} $\Sp(2)$-covariant OPEs. It is unclear to us whether this 
represents a real problem since not all $\pa h^I{}_J$'s appear in 
$T$ anyway.

As in the $d=4$ case the zero-mode saturation rule for the above $d=6$ model 
can be obtained by writing $T$ as a collection of $bc$-type and 
$\beta\ga$-type systems all of weight one and 
using standard methods \cite{Friedan:1985}. 
It is important that $w^2_+$ and $\la^+_2$ are non-trivially related to the 
$\beta'$ and $\ga'$ of the corresponding  weight one $\beta\ga$ system. 
In particular, 
$\ga' = \log \la_2^+$ so that $\de(\ga') = \la_2^+ \de(\la_2^+)$ 
(see \cite{Schiappa:2005} for similar comments). 
One finds that the saturation rule is
\be \label{cansat6}
\lb 0| [ \la_2^+ \prod_{c=1}^3 \tha^c_2 ] \tha^+_1\de(\la^+_1) \tha_2^+\de(\la_2^+) \prod_{c=1}^3 [\tha_1^c \de(\la_1^c)] |0 \rb \neq 0  \,,
\ee
which because of the $\de$-functions is equivalent to 
\be
\lb 0| (\la_2 \ga^m \tha_2 )(\tha_2 \ga_m \tha_2)  |\Omega \rb \neq 0 \,,
\ee
where (as in $d=4,10$) $|\Omega \rb = \prod_{A=1}^5 Y^A |0\rb$ with 
$Y^A = C^{AI}_\al \tha_I^\al \de(C^{AI}_\al \la_I^\al)$ for certain 
$C^{AI}_\al$'s. The operators $Y^A$ each carry $(\la,\tha)$ charge $(-1,+1)$ 
and the above saturation rule is consistent with the background charges in 
$T$. 

Note that the above saturation rule is not $\Sp(2)$-covariant 
might be related to the difficulties in constructing all the $\pa h^I{}_J$'s  
at the quantum level. 

To summarise: in this section we analysed the conformal field theory for the 
$d=6$ pure-spinor superstring and showed that it has vanishing central charge 
and is such that the Lorentz current algebra has level one. 
We also obtained the zero-mode saturation rule. As in $d=4$, many 
open problems remain.

\section{The $d=2$ pure-spinor superstring}
\label{d2}

An even simpler (albeit somewhat degenerate) case occurs in $d=2$. We briefly 
consider the case of $\cN=(2,0)$ 
supersymmetry\footnote{We should point out that because of the peculiar 
nature of two dimensions the model and the results in this section 
should be taken with a grain of salt. In particular, some equations and 
normalisations differ from the corresponding ones in $d=4,6,10$ and this may 
be an indication that some aspects of the model require modification.}.  
The left--moving (holomorphic) ``matter'' worldsheet fields are taken to be 
$(x^{m}, \theta_I,p^I)$, where $I=1,2$ (the spinor indexes only take 
one value so we do not write them explicitly). The Grassmann-odd fields 
$\theta_I$ are Majorana-Weyl spinors and $p^I$ are their conjugate momenta. 
The R-symmetry group is $\SO(2)$. 
We take the  world--sheet ghost fields to be bosonic Majorana-Weyl spinors 
$\la_I$, satisfying the constraint
\be \label{2d-pure-spinor}
\de^{IJ} \lambda_I \gamma^{m} \lambda_J  = 0\,, \qquad I,J=1,2 \,.
\ee
In the $\U(1)$ basis the pure-spinor constraint reads 
$\sum_I (\la_I^+)^2 = 0$. 
Assuming the $\la^+_I$'s are complex we can eliminate $\la^+_2$ and 
gauge-fix $w^2_+=0$.  Using the 
short-hand notation $\la^+_1 = \la^+$ and $w^+=w_1^+$, 
one finds that in the $\U(1)$ basis, 
the Lorentz-invariant quantities are:
\be
N^1{}_1 = \half w_+ \la^+ \,,
\ee
and (using a convenient normalisation)
\be
\pa h = w_+ \la^+ + \frac{\pa \la^+}{\la^+} \,.
\ee
From the above expressions it follows that $N^{mn}$ forms an $\SO(2)$ 
current algebra 
with level $k=-1$. Thus $L^{mn} = N^{mn} + M^{mn}$ has level $k=1$ as 
required, since $M^{mn} = -\half p^I \ga^{mn} \tha_I$ has level $k=2$. 
Furthermore, $\pa h$ has no singularities 
with $N^{mn}$, and satisfies
\be
h(y) h(z) \sim \log \left( y - z \right), \quad \partial h(y)
\lambda_I(z) \sim  \frac{1}{y-z} \lambda_I(z)\,.
\ee
(Note that these OPEs differ slightly from the corresponding ones  
in $d=4,6,10$.)
Finally, using normal-ordering rearrangements, one finds
\be \label{2dTNh}
T_{N,\pa h} 
=  {\ts \frac{1}{4}} N_{mn} N^{mn} + \half \left
( \partial h \right)^{2}  = 
( (w_+\la^+) + \half \frac{\pa \la^+}{\la^+} ) \frac{\pa \la^+}{\la^+} = 
-\beta' \, \pa \ga'\,,
\ee
where we have defined $\beta' = - ((w_+\la^+) 
+ \half \frac{\pa \la^+}{\la^+} )$ and $\ga' = \log \la^+$. Here $\beta',\ga'$ satisfies the usual OPEs of a weight 
one $\beta\ga$-system. The central charge of (\ref{2dTNh}) 
is $c=+2$. Thus the total 
central charge vanishes since  $T_{\rm mat} = -\half \pa x^m \pa x_m  
- \de^{IJ} p_I \pa \tha_J$ has $c=2-4=-2$.

Using the above results and $\de(\ga') = \la^+ \de(\la^+)$ one finds 
that the saturation rule becomes 
\be \label{2dcansat}
\lb 0| \ep^{IJ}\tha^+_I \tha^+_J \la^+ \de(\la^+)) |0 \rb \neq 0  \,.
\ee
which is equivalent to
\be \label{satint}
\lb 0 | \ep^{IJ} \la_I \tha_J |\Omega \rb\neq 0  \,,
\ee
where $|\Omega\rb = Y |0\rb $ and 
$Y = C^I \tha_I \, \de(C^I \la_I )$ for some $C^I$.

\setcounter{equation}{0}
\section{Relation to RNS?} \label{RNS}

In this section we discuss the relation of the new lower-dimensional 
pure-spinor models to compactifications of the RNS superstring.
But before turning to the new models it is useful to recall 
what is known about the map between the $d=10$ RNS and pure-spinor 
superstrings~\cite{Berkovits:2001a} (see also \cite{Aisaka:2003}). 
We should stress that the knowledge of this map is not yet at the level of 
a rigorous proof of the equivalence between RNS and the pure-spinor 
superstring. To fix notation we collect the bosonisation formul\ae{} 
for the RNS ghost variables $(\beta,\!\ga,b,c)$
\be \label{RNSbos}
\beta = \pa \xi e^{-\phi}, \quad \ga = \eta e^{\phi}, \quad
\xi = e^{\chi}, \quad 
\eta = e^{-\chi}, \quad
c = e^{\si}, \quad b = e^{-\si} \,,
\ee
as well as those for the RNS worldsheet fermions, $\Psi^m$ 
(here $a=1,\ldots,5$)
\be
\psi_a \equiv \Psi^{a} - i\Psi^{a+5} = e^{+ \tau^a} \,, \qquad
\psi^a \equiv \Psi^{a} + i\Psi^{a+5} = e^{- \tau^a}  \,.
\ee
To relate the RNS and pure-spinor superstrings, 
the first step is to change variables from the RNS variables to the 
(GSO projected) variables introduced 
in~\cite{Berkovits:2001a,Berkovits:1999a} by Berkovits.  
In terms of these variables, a $\U(5)$ subgroup of the 
(Wick-rotated) $\SO(10)$ super-Poincar\'e symmetry is 
manifest~\cite{Berkovits:1999a}. 
In addition to the $x^m$'s, which are left unchanged, 
the new variables comprise the 12 Grassmann-odd variables 
\be
\ba{lll} 
&\theta^a = e^{\phi/2 -\tau^a +\sum_b \tau^b/2} \,, & \qquad
\theta^+   =  e^{\si + \chi -3\phi/2 -\sum_a \tau^a/2}\,, \non \\
& p_a = e^{-\phi/2 + \tau^a - \sum_b \tau^b/2}\,, & \qquad
p_+ = e^{-\si-\xi + 3\phi/2 + \sum_a \tau^a/2} \,,
\ea
\ee
as well as the two Grassmann-even ones
\be \label{u5even}
s = \si -{\ts \frac{3}{2}}\phi -\half \sum_{a=1}^{5} \tau^a \,, \qquad
t = -\chi +{\ts \frac{3}{2}}\phi +\half \sum_{a=1}^{5} \tau^a \,.
\ee
The next step is to add to the above variables the ten ``topological'' 
quartets $(p^{ab},\tha_{ab},v^{ab},u_{ab})$ (here $p_{ab}=-p_{ba}$ etc.).
This is a sum of ten $bc$- and ten $\beta\ga$-type systems 
all with weight one and consequently each quartet has central charge $c=0$.   
After the addition of these quartets the field content 
is exactly that of the pure-spinor superstring. The Grassmann-odd variables 
$(\tha^+,\tha^a,\tha_{ab})$ span a 16-dimensional spinor $\tha^\al$ and 
$(\la^+,\la_{ab})=(e^s,u_{ab})$ make up the eleven components of a 
pure spinor, $\la^\al$.

In terms of the pure-spinor variables (and after adding the ten quartets)
 the RNS stress tensor 
becomes (suppressing the $x^m$ piece)
\be \label{RNSstress}
T = \pa s \pa t + \pa^2 s + \half v^{ab} \pa u_{ab} 
- p_+ \pa \tha^+ - p_a \pa \tha^a - \half p^{ab} \pa \tha_{ab}  \,.
\ee
The following OPEs are non-vanishing  
(our conventions are as in \cite{Schiappa:2005})
\bea \label{RNSOPEs}
t(y)\, s(z) \sim \log(y-z) \,, && v_{ab}(y)\, u^{cd}(z) 
\sim -\frac{\de^{cd}_{ab}}{y-z}\,, \non\\
p_{+}(y)\, \tha^+(z) \sim \frac{1}{y-z} \,, &&
p_{a}(y)\, \tha^{b}(z) \sim \frac{\de^b_a}{y-z} \,,\non \\
p_{ab}(y)\, \tha^{cd}(z) \sim \frac{\de^{cd}_{ab}}{y-z}\,.
\eea 

It is also useful to recall the discussion of the zero-mode saturation rule 
presented in~\cite{Schiappa:2005} (see \cite{Schiappa:2005} for further 
details). By redefining the $s$, $t$ variables according to 
\be 
s = \half(\chihat - \phihat) \,, \qquad t = \chihat + \phihat \,,
\ee
one finds in the $s$, $t$ sector
\be \label{ststress}
T_{s,t} = \half \pa \chihat \pa \chihat + \half \pa^2 \chihat 
- \half \pa\phihat\pa \phihat - \half \pa^2 \phihat\,,
\ee
which one recognises (see e.g.~\cite{Polchinski:1998}) as
a bosonised $\behat\gahat$-system with weight one and therefore 
$T_{s,t}=-\behat\pa\gahat$. 
Note that this is {\em not} the same as the usual RNS $\beta\ga$-system. 
From this result one sees that the above stress tensor (\ref{RNSstress}) 
is simply the sum of sixteen $bc$-type systems and eleven $\beta\ga$-type 
systems, all of weight one. Using standard methods~\cite{Friedan:1985} 
one then finds  the following zero-mode 
saturation rule (in the small Hilbert space with respect to the eleven 
$\beta\ga$-systems) 
\be \label{cansat}
\lb \ep_{abcde}\tha^a\tha^b\tha^c\tha^d\tha^e
[\tha^{+} \prod_{ab=1}^{10} \tha_{ab} ]
[\de(\gahat) \prod_{ab=1}^{10} \de(u_{ab}) ] \rb \neq 0  \,.
\ee
The relations between $s$, $u_{ab}$ and the eleven components 
of the pure spinor are $\la_{ab} = u_{ab}$ and 
$\la^+ = e^s = \gahat^{-1/2}$. 
Using these relations one can write the 
above saturation rule as~\cite{Schiappa:2005} 
(note that $\de(\gahat) \propto (\la^+)^3 \de(\la^+)$)
\be \label{prelsat}
\lb (\la^+)^3\ep_{abcde}\tha^a\tha^b\tha^c\tha^d\tha^e
[\tha^{+} \prod_{ab=1}^{10} \tha_{ab} ]
[\de(\la^+) \prod_{ab=1}^{10} \de(\la_{ab}) ] \rb \neq 0  \,,
\ee
which was argued in~\cite{Schiappa:2005} to be equivalent to
\be \label{endsat}
\lb (\la \ga^m \tha) (\la \ga^n \tha) (\la \ga^p \tha) (\tha \ga_{mnp} \tha) 
\prod_{I=1}^{11} Y_{C^I} \rb \neq 0  \,,
\ee
where $Y_{C^I} = C^I_{\al} \tha^{\al} \de(C^I_{\beta} \la^{\beta})$, and
$C^I_{\al}$ are certain non-covariant constant spinors. The result 
(\ref{endsat}) precisely coincides with the saturation rule proposed 
in~\cite{Berkovits:2004a} (see also~\cite{Chesterman:2004}). 

Let us now turn to the lower-dimensional pure-spinor models. 
We will argue that these models correspond to the compactification-independent 
sector of RNS compactified on Calabi-Yau manifolds. 

Let us start by discussing the $d=4$ case. 
Since the pure-spinor models discussed in 
previous sections only preserve a lower-dimensional supersymmetry one is not 
restricted to make the same change of variables in all dimensions. 
If we keep $\tha^+$ and $\tha^a$ ($a=1,2$) as above but instead 
of $\tha^a$ ($a=3,4,5$) use\footnote{Note that 
$\Upsilon^i = c e^{-\phi} \psi^{i+2}$ in terms of the RNS variables.} 
$\Upsilon^i = e^s\tha^{i+2}$ ($i=1,\ldots,3$) 
and their conjugates $r_i = e^{-s} p_{i+2}$, we find that the RNS stress 
tensor becomes
\be
T = \pa s \pa \tilde{t} - \half \pa^2 s - p_+ \pa \tha^+ 
- \sum_{a=1}^{2} p_a \pa \tha^a 
- \sum_{i=1}^{3} r_i \, \pa \Upsilon^i \,,
\ee
where $\tilde{t} = t - \frac{3}{2} s - \frac{3}{2} \phi 
- \frac{3}{2} \sum_{a=1}^{2} \tau^a - \frac{1}{2} \sum_{i=1}^3 \tau^{i+2}$ 
and, importantly, compared to (\ref{RNSstress}) the coefficient 
in front of $\pa^2 s$ has changed. 
The non-vanishing OPEs between the variables are the same as 
in (\ref{RNSOPEs}) with the replacements 
$t \rar  \tilde{t}$, $p_{i+2} \rar r_i$ and $\tha^{i+2} \rar \Ups^i$. 

By redefining the $s$, $\tilde{t}$ variables 
according to 
\be \label{st}
s = -\chihat + \phihat \,, \qquad \tilde{t} = \half(\chihat + \phihat) \,,
\ee
one finds the same stress tensor as in (\ref{ststress}), i.e.~$T_{s,\tilde{t}}
=-\behat\pa\gahat$, the difference being that now $\gahat = e^s$ instead of 
$\gahat = e^{-2s}$.
If we now add one  ``topological'' quartet 
$(p^{ab},\tha_{ab},v^{ab},u_{ab})$ (here $a,b=1,2$ and $p_{ab}=-p_{ba}$ etc.) 
then the field content and stress tensor becomes exactly that of 
the pure-spinor superstring plus a decoupled ``internal'' 
$(r_i,\Ups^i)$ sector. 
This can be seen as follows: $(\tha^+,\tha^a,\tha_{ab})$ span a 
four-dimensional Grassmann-odd spinor $\tha^\al$, and 
$(\la^+,\la_{ab})=(e^s,u_{ab})$ make up the pure spinor (cf. section \ref{d4}).
Note that the map we constructed is really only a local map in the sense that 
it maps RNS into one of the two patches in (\ref{lapar}). This needs to 
be better understood.

One might wonder what is so special about the above change of variables. 
One can of course also make other changes of variables (like the one used 
in the $d=10$ case). The motivation for our change of variables is 
that it is such that four-dimensional super-Poincar\'e covariance is 
obtained at the end (since the end result agrees with the pure-spinor 
model of section \ref{d4}).

Note that the central charge is zero  for the $d=4$ (pure-spinor) 
and $d=6$ (internal $(r_i,\Ups^i)$) pieces separately. 
Thus the inconsistency which should arise if one drops 
the internal piece can not be seen in the central charge calculation.

Note also that the variables in the internal sector, $(r_i,\Ups^i)$, 
transforms as triplets under $\SU(3)$. Since we only have manifest $\cN=1$ 
supersymmetry in the $d=4$ system, it should be possible to 
replace the internal directions with any Calabi-Yau manifold, not just flat 
space as above. The three complex bosonic coordinates in the internal 
directions, $y^i$ say, 
together with the $\Ups^i$ seem to make up a super-Calabi-Yau structure 
since they can be combined into the superfields $Y^i = y^i + \al \Ups^i$, 
where $\al^2=0$. 

One can also discuss the zero-mode saturation rule along the lines of 
the $d=10$ discussion. Since we are dealing with $bc$ and $\beta\ga$ 
systems of weight one (in particular $\gahat = e^s = \la^+$)   
one obtains
\be
 \lb [\prod_{a=1}^2 \tha^a]  
[\tha^+ \de(\la^+) \tha_{ab} \de(\la_{ab})] \rb \neq 0  \,,
\ee
which agrees with the result found in section \ref{d4}. In the internal directions one finds  $\lb \prod_{i=1}^3 \Ups^i \rb \neq 0 $.

To discuss the $d=6$ model one just changes the ranges of the $a$ and $i$ 
indexes. One finds that the RNS stress tensor becomes
\be \label{d6RNSstress}
T = \pa s \pa \tilde{t} - p_+ \pa \tha^+ 
- \sum_{a=1}^{3} p_a \pa \tha^a 
- \sum_{i=1}^{2} r_i \, \pa \Upsilon^i \,,
\ee
where now $\tilde{t} = t -  s - \phi - \sum_{a=1}^{3} \tau^a $. 
Since there is now no $\pa^2 s$ term in $T$ one finds that $\gahat = s$. 

Denoting $(\tha^+,\tha^a)$ by $\tha_2^\al$ and $\la_2^+ = e^s$
and adding the four quartets $(p^1_\al,\tha_1^\al,w^1_\al,\la^\al_1)$ one 
finds exactly the field content of the $d=6$ pure-spinor model discussed in 
section~\ref{d6}: $\tha_1^\al$ and $\tha_2^\al$ span $\tha^\al_I$ and 
 $\la_2^+$ and $\la^\al_1$ make up the pure spinor $\la^\al_I$. 

The saturation rule follows as in the $d=4,10$ cases (in particular 
one needs to use $\gahat = \log \la_2^+$)
\be
 \lb [\la_2^+ \prod_{a=1}^3 \tha_2^a]   
[\tha_2^+ \de(\la_2^+) \tha_1^+ \de(\la_1^+) 
\prod_{a=1}^3 \tha_1^a \de(\la_1^a)] \rb \neq 0  \,,
\ee
which agrees with the result in section \ref{d6}.

Finally to discuss the $d=2$ case we make the change of variables to
 $\tha^+$ and $\tha^a$ ($a=1$) as well as $\Upsilon^i = e^{s/2} \tha^{i+1}$ 
($i=1,\ldots,4$) and their conjugates $r_i = e^{-s/2} p_{i+1}$. 
(The fact that this change of variables is slightly different from the $d=4,6$ 
cases again highlights the peculiar nature of the $d=2$ model.)
Under the change of variables the RNS stress 
tensor becomes exactly the same as in the $d=6$ case (\ref{d6RNSstress}), 
except that the ranges of $a$ and $i$ are different (and the definition 
of $\tilde{t}$ is slightly different). If we use the notation 
$\tha_I = (\tha^+,\tha^a)$ and $\la_1 = e^s$ one finds that the field content 
and stress tensor are exactly the same as for the $d=2$ pure-spinor model 
discussed in section \ref{d2}. 

As in $d=6$ we have $\gahat = s =\log \la^1$ so the saturation rule becomes
\be
 \lb [\la_1 \tha_2]   
[\tha_1 \de(\la_1)] \rb \neq 0 \,,
\ee
which agrees with the result in section \ref{d2}.

A natural question to ask is the following. If the pure-spinor models discussed in this paper are to be thought of as compactifications of the RNS superstring 
it should also be possible to understand them more directly as 
compactifications of the $d=10$ pure-spinor superstring. 
Let's see how this might work for the $d=4,6$ models. If we write the $d=10$ variables in $\U(5)$ representations and decompose them under the subgroup  
$\U(2){\times}\U(3)  \subset \U(5)$
we find 
\be
\ba{llll}
&(\tha_{ab},\la_{ab})&:& \quad10 \rar (1,1)\oplus(1,3)\oplus(2,3) \,, \non \\
&(\tha^a)  &:& \quad 5\rar(2,1)\oplus(1,3) \,, \\
&(\tha^+,\la^+) &:& \quad1\rar (1,1) \,. \non
\ea
\ee
Looking at the variables that are singlets under $\U(3)$ we see that these  
variables make up the $\U(2)$ field content $\la^+,\la_{ab}$ and 
$\tha^+,\tha^a,\tha_{ab}$ i.e. exactly the same field content as in 
the $d=4$ pure-spinor model. Assuming that the quartets in the 
representations $(1,3)$ and $(2,3)$ can be removed, we are left with 
(in addition to the above variables) one $\SU(3)$ triplet 
arising from $\tha^a$. Thus this naive counting leads to the same representation content as the $d=4$ pure-spinor model. The counting for the $d=6$ works 
in a similar way by interchanging the roles of $\U(2)$ and $\U(3)$ and  
assuming that the $(2,3)$ quartets can be removed. The $\U(2)$ singlets make up the representation content of the $d=6$ pure-spinor model and there is in addition an $\U(2)$ doublet coming from $\tha^a$.  
It would be interesting to make the rules for compactification more precise.

To summarise: we have presented some evidence in favour of a relation 
between compactifications of the RNS superstring and the lower-dimensional 
pure-spinor models. At this stage many of the results are heuristic and it 
would be nice if one could understand the relation to the  
RNS superstring better.

\setcounter{equation}{0}
\section{Relation to hybrid superstrings?} \label{hybrid}

There exists (quantum) formulations of superstring theory compactified on 
the Calabi-Yau manifolds $CY_{l}$ ($l=2,3,4$) with manifest $SO(10-2l)$ 
super-Poincar\'e symmetry in the non-compact dimensions. These 
formulations are referred to as hybrid 
superstrings~\cite{Berkovits:1994a,Berkovits:1994b}
and can be obtained via a change of variables from the RNS superstring. 
Such formulations exist for $d=4$ ~\cite{Berkovits:1994a},
$d=6$~\cite{Berkovits:1994b,Berkovits:1999b} (see also~\cite{Berkovits:1999d}) 
and $d=2$~\cite{Berkovits:2001f}. For reviews, 
see e.g.~\cite{Berkovits:1995a, Berkovits:1996b,Berkovits:1999d}.

The hybrid description has been most extensively developed for $d=4$ where 
vertex operators for massless \cite{Berkovits:1994a} 
(see also~\cite{Berkovits:1996b}) and 
the first massive states \cite{Berkovits:1997} have been studied. 
Scattering amplitudes for massless modes have also been investigated at 
tree-level \cite{Berkovits:1996a} and at one loop \cite{Berkovits:2001e}. 
Effective actions have been studied in \cite{Berkovits:1995b}. 

In $d=4$ the hybrid formulation is obtained via a change of variables 
starting from the RNS superstring in bosonised form. The part of 
the stress tensor in the $d=4$ hybrid superstring containing 
the $d=4$ modes is (see e.g.~\cite{Berkovits:1994b}) 
\be
T = -\half \pa x^m \pa x_m - p_a \pa \tha^a 
- \bp_{\dot{a}} \pa \btha^{\dot{a}} - \frac{1}{2} \pa \rho \pa \rho
\ee
where the chiral scalar $\rho$ has the OPE 
$\rho(y) \rho(z) \sim - \log(y-z)$. In the hybrid model
there is also a $\U(1)$ current, whose $d=4$ part is $J = \pa \rho$. 
After ``twisting'' $T \rar T  + \half \pa J$ we find 
$T_{\rho} = - \frac{1}{2} \pa \rho \pa \rho + \frac{1}{2} \pa^2 \rho$. 

Comparing the (twisted) stress tensor in the hybrid superstring to the one 
in the pure-spinor superstring written in terms of 
$\pa h$ and $N^{mn}$ (\ref{4dTmat}), (\ref{covst4N}) we see that 
they almost agree provided that we identify $\pa \rho = \pa h$. 
There is still a difference though since the pure spinor expression also 
contains the additional piece $T = -\frac{1}{8} N_{mn} N^{mn}$. This extra 
piece has  $c=0$ and is in some sense ``topological'', similar to the 
quartets which were added to the RNS superstring to obtain the pure-spinor 
models (see the previous section).

One can also compare the vertex operators in the pure-spinor and 
hybrid formalisms. In \cite{Grassi:2005} we studied the integrated 
vertex operator for the massless modes obtained from the naive BRST operator 
$Q = \oint \la^\al d_\al$. In the Weyl basis it take the form (similar to 
the $d=10$ result) 
\be
V = \oint [ \pa x^m A_m + \pa \tha^a  A_a + \pa \btha^{\dot{a}} \bA_{\dot{a}}
+ d_a W^a + \bd_{\dal} \bW^{\dal} + \half N^{mn} F_{mn} ]  
\ee
Here the superfields 
$A_m$, $A_a$, $\bA_{\dot{a}}$,  $W^a$, $ \bW^{\dal}$ and $F_{mn}$ depend 
on $x^m$ and $\tha^\al$ and can all 
be expressed in terms of one scalar superfield (see \cite{Grassi:2005} 
for details). If one compares the resulting expression to the result in the 
hybrid superstring (see e.g.~\cite{Berkovits:1994b}, section 6.1) 
one finds perfect agreement except for the fact that in the hybrid case 
the $F_{mn} N^{mn}$ piece is missing. 
 This is of course consistent with our observation that the 
difference between the hybrid and pure-spinor models seems to be 
the addition of a  $N^{mn}$, $c=0$ sector.

Roughly, the suggested relation between the pure-spinor and hybrid models
 can be summarised as in the figure below 
\begin{figure}[h]
\centering
\includegraphics{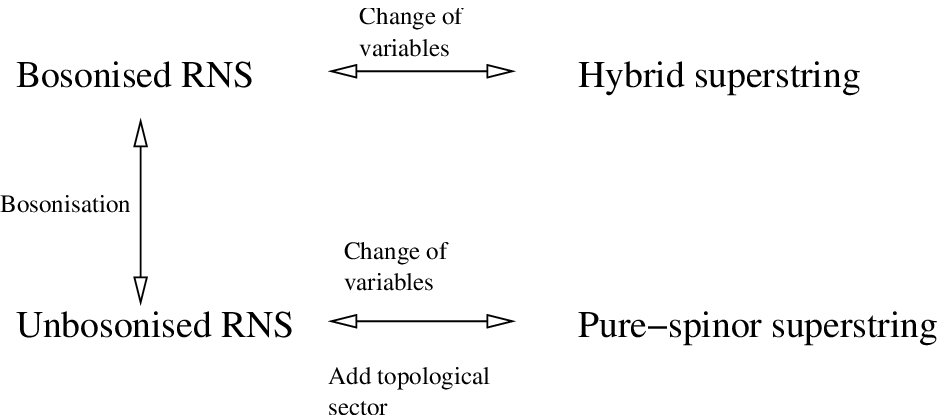}
\center{\small {\sf Figure 1:} 
Schematic overview of suggested relation \\ between  
hybrid and pure-spinor models in $d=4$.}
\end{figure}

The role of the additional $N^{mn}$ in the pure-spinor model as compared 
to the hybrid model is that one gets an interpretation in terms of 
pure spinors. In the pure-spinor superstring one can in many cases write 
things in terms of $\pa h$ and $N^{mn}$, but one also has the pure spinor 
interpretation. Another comparison one can make is to compare the saturation 
rules in the two models. 
The saturation rule for the $d=4$ hybrid model is 
(see e.g.~\cite{Berkovits:1996b})
\be
0 \neq \langle e^{\rho} \rangle = \langle e^{ h } \rangle  \,,
\ee
where we used our identification of $\rho$ and $h$. 
 On the other hand, in the 
pure-spinor formulation one has 
\be
0 \neq \langle \de(\la^+) \de(\la_{ab}) \rangle 
= \langle e^{ h + \log(\la^+) } 
\rangle \,,
\ee
where we have used that (by bosonising) $\de(\la^+) = e^{-\phi^+}$ where 
$w_+ \la^+  = -\pa \phi^+$ and similarly for $\de(\la_{ab})$ and compared 
with our earlier expression for $\pa h$ (\ref{4dpah}). Thus we see 
that although both saturation rules are consistent with the background 
charge of $\pa h$ they are not the same. 
The difference might be related to a choice of large vs. small Hilbert 
spaces. 
Another interpretation might be that maybe in the hybrid model  
$\log(\la^+)$ has somehow been absorbed into the internal Calabi-Yau piece 
$H_C$ (the complete saturation rule in the hybrid model 
is $0 \neq \langle e^{\rho + H_C} \rangle $). 

One could also investigate if the $\cN=2,4$ superconformal generators of the 
hybrid superstring can be written in pure-spinor language, as well as 
 investigate the connection between the $\la$ that appears in the twistor 
formulation (see e.g.~\cite{Berkovits:1996b}) and the pure-spinor $\la$.

Next we turn to the $d=6$ model. In the $d=6$ hybrid formulation the part of 
the (untwisted) stress tensor containing the $d=6$ modes 
is \cite{Berkovits:1994b}
\be
T = -\half \pa x^m \pa x_m - p^2_\al \pa \tha_2^\al 
- \frac{1}{2} \pa \rho \pa \rho - \frac{1}{2} \pa \si \pa \si 
- \pa^2(\rho + i \si) \,,
\ee
where $\rho$, $\si$ have the OPEs 
$\rho(y) \rho(z) \sim - \log(y-z)$ and $\si(y) \si(z) \sim - \log(y-z)$. 
In the hybrid model
there is also a $\U(1)$ current, whose $d=6$ part 
is $J = -\pa (\rho + i \si)$. 

In the $d=6$ pure-spinor superstring the stress tensor in the ghost sector 
contains (see section \ref{d6}) $T_{u,v} = \pa u \pa v - \pa^2v$.  
 By redefining 
$u = - \half(\mu - i \nu)$ and $v = \mu + i \nu$ one finds 
\be 
T_{\mu,\nu} = -\half \pa \mu \pa \nu - \half \pa \nu \pa \nu -  \pa^2(\mu +i\nu) \,. 
\ee
Comparing this to the $(\rho,\si)$ part of the above expression 
in the hybrid superstring one finds 
agreement provided one identifies $\mu \rar \rho$ and $\nu \rar \si$. 
Furthermore, using the $\pa u \pa v$ OPE one finds 
$\mu(y) \mu(z) \sim -\frac{1}{(y-z)^2}$ and 
$\nu(y) \nu(z) \sim -\frac{1}{(y-z)^2}$, so also the OPEs agree 
with the results in \cite{Berkovits:1994b}. 
Furthermore, under the identification 
$e^{-\rho-i \si} = e^{-v} = \la^+_2/\la_1^+$. Note that in contrast to the 
$d=4$ case,  no twisting was needed to get agreement. We do not understand 
the reason for this difference.

Although the above pieces agree, in the full stress tensor 
there is a difference though  
since in the pure spinor case one has an additional $c=0$ piece: 
$T = -\pa p^1_\al \pa \tha^\al_1 - \frac{1}{12} N_{mn}N^{mn} 
+\half \pa h \pa h - \pa^2 h$ . 
This situation is similar to the $d=4$ case discussed above. 
Note that if one writes 
$N^{mn} = \half \ka \ga^{mn} \ze$ and $\pa h = \half \ka \ze$ in terms of two 
canonically conjugate Grassmann-even Weyl spinors $\ka_\al$ and $\ze^\al$, 
then $N^{mn}$ and $\pa h$ satisfy the same OPEs as in section~\ref{d4}. 
Furthermore, it can be shown that 
$ - \frac{1}{12} N_{mn}N^{mn} 
+ \half \pa h \pa h - \pa^2 h = \ka_\al \pa \ze^\al$. This result makes 
contact with the comment in footnote 3 of \cite{Berkovits:2000a}. 
One also sees explicitly that the additional $c=0$ piece present 
in the pure-spinor superstring, as compared to the hybrid model, 
comprise four ``topological'' quartets 
$(p^1_\al,\tha^\al_1,\ka_\al,\ze^\al)$.  

Furthermore, if one uses the solution to the pure-spinor constraint, 
the naive pure-spinor BRST current can be written as 
$\la^\al_I d_\al^I 
= \la^\al_1 ( d_\al^1 + \frac{\la^+_2}{\la^+_1} d_{\al}^2) = 
\la^\al_1 ( d_\al^1 + e^{-v} d_{\al}^2) $. 
This can be compared with the ``harmonic'' constraint introduced 
in \cite{Berkovits:1999b}: 
$d_\al^2 - e^{-\rho - i \si} d_{\al}^1 \approx 0$. 
Using the above relation between $\rho$, $\si$ and $v$ one finds 
$e^{-\rho - i \si} = e^{-v}=\frac{\la^+_2}{\la^+_1}$. After trivially 
interchanging the labels $1\leftrightarrow 2 $  in the harmonic constraint 
and redefining $\tha_1$ and $d^1$ using the symmetry $\tha \rar - \tha$ 
and $d \rar -d$ one finds agreement.

To summarise: in this section we studied the relation between the pure-spinor 
models and the hybrid superstrings. The general structure of the relation 
between these models seems to be that (possibly after ``twisting'') 
the hybrid stress tensor agrees with part of the pure-spinor stress 
tensor. The pure-spinor stress tensor also contains an additional 
$c=0$ piece. Although our results are suggestive, we do not yet have a 
rigorous proof of the equivalence between the hybrid and pure-spinor models.

\setcounter{equation}{0}

\section{Discussion and applications} \label{sDisc}

In this paper we studied various aspects of the lower-dimensional 
pure-spinor superstrings introduced in \cite{Grassi:2005}. 
Actually, referring to these models as pure-spinor superstrings is slightly 
premature since we have not clarified the BRST structure nor have we 
shown how to calculate scattering amplitudes. Pure-spinor conformal field 
theories might be a more appropriate name. However, we presented a  
tentative analysis which suggest a relation of the lower-dimensional 
pure-spinor models  to the hybrid superstrings and to RNS compactified 
on Calabi-Yau manifolds. 
Even if these relations were to turn out not to be true, 
the lower-dimensional models would still be very interesting 
as toy models of the $d=10$ pure-spinor model.

Despite the fact that we do not fully understand all aspects of the models, we 
will now discuss some possible applications of the new models. 
The first application is to curved backgrounds 
of the form $adS_n{\times}S^n$,  and the second application 
is to indicate the possibility of constructing 
 ``pure-spinor M-theories'' in $d=5,7$. In both cases we will be very brief.

\subsection{ Pure-spinor superstring theory in $adS_2{\times}S^2$ 
and $adS_3{\times}S^3$}

One of the most promising aspects of the pure-spinor formalism is that it 
can handle backgrounds with Ramond-Ramond fields turned on. One application 
of obvious interest is to study string theory in $adS_5{\times}S^5$ using the 
pure-spinor formalism.  
It has been shown that in this background the pure-spinor superstring 
sigma-model is quantisable~\cite{Berkovits:2000a}. 
Unfortunately the sigma model is interacting and quite complicated. 
Although some impressive results have been obtained 
(see e.g.~\cite{Berkovits:2000a,Berkovits:2000e,Vallilo:2003,
Berkovits:2004e,Berkovits:2004d}) 
one would like to have a better understanding of the quantum properties 
of the sigma model. 
One approach would be to study simpler lower-dimensional models 
sharing several features with the $d=10$ $adS_5{\times}S^5$ model. 
Such models have been studied in several papers (see 
e.g.~\cite{Berkovits:1999c} for $adS_2{\times}S^2$ models 
and \cite{Berkovits:1999d,Berkovits:1999b,Dolan:1999} 
for $adS_3{\times}S^3$ models). 
As we will now argue, the pure-spinor models introduced in \cite{Grassi:2005}
can also be studied in  $adS_2{\times}S^2$ and  $adS_3{\times}S^3$ 
backgrounds and thus furnishes us with examples of simplified models 
similar to the $adS_5{\times}S^5$ pure-spinor model. 
Of course these models are not really new since they are very closely 
related to the models in~\cite{Berkovits:1999c, Berkovits:1999d,
Berkovits:1999b}. The only difference compared with those models is that since 
the conformal field theories of the models discussed in this paper 
correspond more closely to the $d=10$ pure-spinor model than, say, 
the hybrid models, the models in $adS_2{\times}S^2$ and $adS_3{\times}S^3$ 
will also correspond more closely to the $d=10$ $adS_5{\times}S^5$ 
pure-spinor model. 

The $adS_5{\times}S^5$ pure-spinor sigma model can be written
\bea \label{adSact}
&&\int \frac{1}{2} g_{\ua\ub} J^{\ua} \bar{J}^{\ub}  + 
\frac{3}{4} \de_{\al \tilde{\beta}} J^\al \bar{J}^{\hat{\beta}} +
\frac{1}{4} \de_{\al \tilde{\beta}} J^{\tilde{\beta}} \bar{J}^{\al} \non \\
 &+&\!\!\!\int \frac{1}{2} N_{\uc\ud} \bar{J}^{\uc\ud} 
+ \frac{1}{2} \tilde{N}_{\uc\ud} J^{\uc \ud} 
+ \frac{1}{4} N_{\uc\ud} \tilde{N}^{\uc\ud} + \int w \bar{\pa} \la 
+ \tilde{w} \pa \tla \,,
\eea
where the currents $J^{\ua},J^\al,J^{\tal},J^{\uc\ud}$ (and the $\bar{J}$'s) 
belong to the Lie superalgebra $\mathrm{PSU}(2,2|4)$. In particular, 
$J^{\uc\ud}$ belong to the $\SO(1,5){\times}\SO(6)$ subalgebra.
The terms in the first line of (\ref{adSact}) are similar to the GS action 
constructed in~\cite{Metsaev:1998} and the terms in the second line of 
(\ref{adSact}) describe the couplings to the pure-spinor ghost sector. 

Although we only discussed the lower-dimensional pure-spinor models 
in a flat supergravity background, it is 
straightforward to generalise to the case of a curved background space. 
For the cases of $adS_2{\times}S^2$ and $adS_3{\times}S^3$ the resulting 
models will take the same form as in (\ref{adSact}) except that the currents 
now belong to the Lie superalgebras $\mathrm{PSU}(1,1|2)$ and 
$\mathrm{PSU}(1,1|2){\times}\mathrm{PSU}(1,1|2)$, respectively. 
The reason that the only difference is in the ranges of the (suitably defined) 
indexes is that these 
Lie superalgebras are very similar to $\mathrm{PSU}(2,2|4)$. 
For the lower-dimensional models, the first line in (\ref{adSact}) 
is related to the GS actions constructed 
in~\cite{Pesando:1998} and \cite{Zhou:1999} and 
the second line describes the coupling to the Lorentz currents in the 
ghost sector (we are assuming that there are no couplings to the Lorentz 
scalars).

At one loop it was shown in \cite{Berkovits:1999c} that the terms in 
the first line of (\ref{adSact}) are conformally invariant for 
the $d=4,6,10$ cases.  
The terms in the second line of (\ref{adSact}) were shown to be conformally 
invariant (at one loop) for the $d=10$ case in \cite{Vallilo:2002}. 
In fact that calculation was done for general $SO(2n)$ and it seems 
that if one uses also the result for the level $k=2-n$ the argument works 
also for the $n=2,3$ cases, although we did not check this in detail. 
At higher loops one can use the argument 
in \cite{Berkovits:2004d} which seems to go 
through {\it mutatis mutandis}, although again we 
did not check this in detail. We should also point out that 
conformal invariance to all orders was shown in  \cite{Berkovits:1999b} 
for the hybrid $adS_3{\times}S^3$ model.

One can also investigate the classical flat currents which were constructed 
in~\cite{Vallilo:2003} by generalising the results 
in~\cite{Bena:2003}, as well as study their quantum 
properties~\cite{Berkovits:2004e}. 

It remains to be seen if there are some calculations that can be done that are 
significantly simpler for the lower-dimensional models. We leave this question 
for future work.

\subsection{Pure-spinor M-theory superparticle in $d=5,7$}

In~\cite{Berkovits:2002a} the worldvolume action for the M2-brane 
in $d=11$ was formulated using a pure-spinor formalism. Quantisation 
of the resulting model is complicated, but the superparticle limit 
is well defined and can be studied~\cite{Berkovits:2002a,Anguelova:2004}. 
In particular, in~\cite{Anguelova:2004} covariant scattering amplitudes 
in $d=11$  were calculated using the superparticle formalism.

A natural question to ask is if the M2-brane 
pure-spinor action (or its superparticle limit) can be generalised 
to the other dimensions were (classically) a $\ka$-symmetric GS action 
exists, i.e.~$d=(3),4,5,7$ \cite{Achucarro:1987}. 

Before addressing this question let us briefly recall 
the $d=11$ results. The pure-spinor condition 
is $\la \Ga^M \la =0$, where $\la^A$ is a 32-component spinor and 
$\Ga^M_{AB}$ are the $32{\times}32$-dimensional gamma matrices in $d=11$. 
One can decompose these results in $d=10$ language: the spinor splits as 
$\la^A = (\la^\al,\tla_{\al})$, and writing $M=(m,11)$ the pure-spinor 
condition becomes $\la \ga^m \la +\tla \ga^m\tla=0$ and 
$\tla \ga^{11}\la = \tla \la=0$. 
It can be explicitly checked (see~\cite{Anguelova:2004}) that these 
equations imply that the pure spinor in $d=11$ has 23 independent components. 
This number can be understood as arising from the 11 left-moving and 11 
right-moving pure-spinor degrees of freedom of the $d=10$ 
type IIA pure-spinor superstring plus one extra 
mode whose interpretation was discussed in~\cite{Anguelova:2004}. 

The worldline action for the $d=11$ pure-spinor superparticle 
is~\cite{Berkovits:2002a,Anguelova:2004}
\be \label{MP}
 \int \D \tau \Big(P_M \dot{x}^M - \half P_M P^M + \dot{\tha}^A p_A 
+ \dot{\la}^A w_A \Big) \,,
\ee
and is invariant under $\de w_A = \La_M (\Ga^M \la)_A$. Next we try 
to generalise the $d=11$ model to lower dimensions, focusing on 
the definition of the pure spinor in these dimensions.

In $d=5$ the minimal dimension of a spinor is eight and the gamma matrices 
are $8{\times}8$-dimensional. We take the pure spinor, 
$\la^A$ ($A=1,\ldots ,8)$, to satisfy $\la \Ga^M\la=0$ ($M=0,\ldots,4$). 
Using $d=4$ language, the spinor $\la^A$ can be decomposed 
as $\la^A=(\la^\al,\tla_{\al})$. 
One can decompose further into $\U(2)$ representations: 
$\la^\al = (\la^+,\la^a,\la_{ab})$ as in section~\ref{d4}, and 
$\tla_{\al} = (\tla_+,\tla_a,\tla^{ab})$. The pure-spinor condition 
$\la \Ga^M\la=0$ decomposes into $\la \Ga^m\la + \tla \Ga^m \tla =0$ 
and $\la \Ga^5 \tla=0$. In the $\U(2)$ basis these equations can be written as
\be
\la^+\la^a + \la^{ab}\tla_{b} =0 \,,\qquad 
\tla_+\tla_{a} + \la_{ab}\la^b =0\,, \qquad 
\la^+ \tla_+ + \la^a\tla_a + \half \la_{ab}\tla^{ab}=0\,.
\ee
It is easy to see that these equations eliminate three components from the 
eight-dimensional spinor $\la^A$ (e.g.~$\la^a = -\frac{\tla^{ab}\tla_b}{\la^+}$ and $\tla_+ = -\frac{\la_{ab}\tla^{ab}}{2\la^+}$). 
The number of independent components is therefore five. Note that 
$5=2\cdot 2 +1$ where $2$ is the dimension of the pure spinor 
in $d=4$, so the same counting that worked in $d=11$ also works in $d=5$. 

The $d=7$ case can be treated in an analogous way. We take the pure-spinor 
to be $\la^A_I$ where $A=1,\ldots,8$ and $I=1,2$. 
Using $d=6$ language this spinor decomposes in the following 
way: $\la^A_I = (\la^\al_I,\tla^I_{\al})$. The further splitting 
into $\U(3)$ representations is $\la^\al_I = (\la^+_I,\la^a_I)$ and 
$\tla^I_{\al} = (\tla_+^I,\tla_a^I)$. Using these results, the pure-spinor 
condition $e^{IJ}\la_{I} \Ga^M \la_J = 0$ can be written
\be
\ep^{IJ} \la^+_I \la^a_J + \half \ep_{IJ} \ep^{abc}\tla_b^I \tla_c^J = 0\,, 
\quad
\ep_{IJ} \tla_+^I \tla_a^J + \half \ep^{IJ} \ep_{abc}\la^b_I \la^c_J = 0\,, 
\quad
\la_I^+ \tla^I_+ + \la^a_I \tla^I_a = 0\,.
\ee
It can be shown that these equations eliminate five components from the 
sixteen-dimensional spinor $\la^A_I$ 
(e.g.~$\la_2^a = -\frac{\la^{+}_2 \la^a_1}{\la_1^+} 
- \ep^{abc} \frac{\tla^1_b\tla^2_c}{\la^+_1}$ as well as 
$\tla^1_+ = -\frac{\tla^1_{a}\la_1^{a}}{\la_1^+}$ and 
$\tla^2_+ = -\frac{\tla^2_{a}\la_2^{a}}{\la_2^+}$). Thus the 
pure-spinor contains eleven independent components. Note that 
$11=2\cdot 5 +1$ where $5$ is the dimension of the pure spinor 
in $d=6$, so the same counting that worked in $d=5,11$ works also in $d=7$. 

Just as in $d=11$ one can write down the tree-level saturation rule 
for the $d=5,7$ models. 
In $d=2n+1$ one finds the schematic result 
$\langle 0| \la^{2n-3} \tha^{2n-1} |\Om \rangle \neq 0$ for $n=2,3,5$.
The saturation rule seems to depend only on the number of independent 
components of the pure spinor, since, for instance, 
it has the same schematic form in 
the $d=7$ M-theory case as in the $d=10$ (open) superstring case, and it 
has the same schematic form in the $d=5$ M-theory case as in 
the $d=6$ (open) superstring case. 
The action for the $d=5,7$ ``M-theory'' superparticles takes the same form 
as in (\ref{MP}). 

One could clearly analyse these models further but we will not do so here. 
We also note that the $d=3,4$ cases appear to be subtle and will therefore 
not be discussed here.

\medskip

{\bf Note added:} After this work was completed 
the paper~\cite{Berkovits:2005c} appeared. 
In this work pure-spinor superstrings in $d=4$ are also discussed. 
In particular, it is suggested that the $d=4$ pure-spinor superstring, 
with a particular BRST operator, describes a chiral sector 
of superstring theory compactified on a Calabi-Yau manifold down to $d=4$.

\section*{Acknowledgements}
I would like to thank Pietro Antonio Grassi and Ricardo Schiappa for 
collaborations on related topics and for several discussions. 
I would also like to thank Nathan Berkovits, Martin Cederwall 
and Bengt Nilsson for discussions and comments.

\appendix

\setcounter{equation}{0}
\section{The $\mathrm{U}(n)$ Formalism} \label{un}

It will occasionally be useful to temporarily break $\SO(2n)$ to 
$\U(n) \approx \U(1){\times}\SU(n)$. Under this breaking pattern, the vector 
representation of $\SO(2n)$ decomposes as $2n \rar n \oplus \bar{n}$. 
The components of a $\SO(2n)$ vector $V^m$ are related to the 
components of the two $\U(n)$ 
representations $v^a$, $v_a$, according to 
$v^a = \frac{1}{2}( V^a + i V^{a+n})$ for 
the $n$ and $v_a = \frac{1}{2}(V^a - i V^{a+n})$ for the $\bar{n}$; here 
$a=1,\ldots,n$. Analogous expressions can be derived for a tensor with an
 arbitrary number of vector indexes. 
The following representations for the $\U(n)$ components $(\ga^a)_{\al \bet}$ 
and $(\ga_a)_{\al\bet}$ of the $\SO(2n)$ gamma matrices $\Ga^{m}_{\al\bet}$ 
will be used in this paper. In $d=2$ the $2{\times}2$ matrices 
\begin{equation}
\ba{lcl}
(\ga^1)_{\al\beta} = i \frac{1+\si_3}{2}   \,,  & \quad
&  (\ga_1)_{\al\beta} = -i\frac{1-\si_3}{2} \,.
\ea
\end{equation}
are symmetric. In $d=4$ the $4{\times}4$ matrices 
\begin{equation}
\begin{array}{lcl}
(\ga^1)_{\al\beta} = i\frac{1+\si_3}{2} \otimes \si_1  \,, & \quad 
&  (\ga_1)_{\al\beta} = -i\frac{1-\si_3}{2}\otimes \si_1 \,, 
\\
(\ga^2)_{\al\beta} = -i \si_1\otimes \frac{1+\si_3}{2} \,, &\quad 
&  (\ga_2)_{\al\beta} = -i\si_1\otimes \frac{1-\si_3}{2} \,.
\end{array}
\end{equation}
are symmetric. Finally, in $d=6$ the $8{\times}8$ matrices 
\begin{equation}
\begin{array}{lcl}
(\ga^1)_{\al\beta} = i\frac{1+\si_3}{2} \otimes \si_1\otimes \si_2 \,, 
& \quad &  (\ga_1)_{\al\beta} = 
-i\frac{1-\si_3}{2}\otimes \si_1 \otimes \si_2  \,, 
\\
(\ga^2)_{\al\beta} = -i\si_1\otimes \frac{1+\si_3}{2}\otimes  \si_2 \,, 
&\quad &  (\ga_2)_{\al\beta} = 
-i\si_1\otimes \frac{1-\si_3}{2} \otimes \si_2 \,, 
\\
(\ga^3)_{\al\beta} = -i\si_1\otimes \si_2 \otimes \frac{1+\si_3}{2}\,, 
& \quad&  (\ga_3)_{\al\beta} = 
-i\si_1\otimes \si_2\otimes \frac{1-\si_3}{2} \,.
\end{array}
\end{equation}
are antisymmetric.
 
Indexes are raised and lowered with $\mathcal{C}^{\al\bet} = \si_2$ ($d=2$), 
$\mathcal{C}^{\al\bet} = \si_2\otimes \si_1$ ($d=4$) and 
$\mathcal{C}^{\al\bet} = \si_2\otimes \si_1\otimes \si_2$ ($d=6$) 
and its inverse $\mathcal{C}_{\al\bet}$, according to the rule $T^{\al\bet} 
= \mathcal{C} ^{\al \de}T_{\de \rho} \mathcal{C}^{\rho \bet}$.  

The above matrices satisfy $\{\ga^a,\ga_b\}=\de^a_b$. 
From this result it follows that the corresponding $\Ga^m$'s 
satisfy $\{\Ga^m,\Ga^n\} = 2 \eta^{mn}$. 

In $d=6$ (as in $d=10$) the restriction of $\Ga^m_{\al\bet}$ 
to the Weyl subspace (action on Weyl spinors) will be denoted by 
$\ga^m_{\al\bet}$.  

A spinor of $\SO(2n)$ is conveniently represented as the direct product of 
$n$ $\SO(2)$ spinors. Denoting the $\SO(2)$ spinor {\tiny $ 
\left( \!\!\! \ba{c} 1 \\ 0\ea \!\!\! \right)$ } by $+$ and  
{\tiny $\left( \!\!\! \ba{c} 0 \\ 1\ea \!\!\! \right)$ } by $-$, 
$\SO(2n)$ spinors are naturally labelled by a composite index 
$(\pm,\ldots,\pm)$, where all possible choices are allowed. 

The above $\ga$ matrices act on this basis in the natural way. 
Our conventions for the chirality matrix are such that spinors with an 
odd (even) number of $+$'s are Weyl (anti--Weyl) spinors. 
The difference between the number of $+$'s and $-$'s divided by 2 is the 
$\U(1)$ quantum number.

The following notation is used for $\SU(d)$ components of a spinor $\la^\al$.  
In $d=2$ the $+$ (Weyl) component is denoted $\la^+$. 
In $d=4$ the $++$ component is denoted $\la^+$, the components with one 
$+$ are denoted $\la^a$ and the $--$ component is denoted by $\la_{ab}
=-\la_{ba}$ ($a,b=1,2$). We also use the notation $\bla^{\dal}$ for 
$\{ \la^{+}, \half\ep^{ab}\la_{ab}\}$ . Note that $\la^a$ is a Weyl-spinor, 
whereas $\bla^{\dal}$ is an anti-Weyl spinor.
In $d=6$ the $+++$ component is denoted $\la^+$, and 
the components with one $+$ are denoted $\la^a$ ($a=1,2,3$). 
These components span a Weyl-spinor.

\begingroup\raggedright\endgroup

\end{document}